**Title**: Ultrafast Faraday Rotation Probe of Chiral Phonon-Polaritons in LiNbO$_3$

**Authors**: Megan F. Biggs[1], Sin-hang (Enoch) Ho[1], Aldair Alejandro, Matthew Lutz, Clayton D. Moss, Jeremy A. Johnson[2]

Affiliations: Department of Chemistry and Biochemistry, Brigham Young University, Provo, UT 84602, USA

[1]Contributed equally
[2]Electronic Mail: jjohnson@chem.byu.edu

**Abstract**

Time reversal symmetry breaking motion of chiral phonon-polaritons in LiNbO$_3$ is probed via the ultrafast Faraday effect. By combining a pair of perpendicularly polarized THz pulses with the right relative delay, we create a chiral THz driving field to excite chiral phonon-polaritons. The chiral atomic motion combines with the inverse Faraday effect from the circularly polarized THz pump to induce a magnetic moment field in the nonmagnetic material, LiNbO$_3$. We attempt to quantify the strength of the magnetic field with Faraday rotation probe measurements. The direction of the Faraday signal flips when the input THz pulse is changed from left- to right-circular polarization, and we estimate a strong induced magnetic field strength of ~11 Tesla based on the Faraday rotation.

**Introduction**

Motion of charged particles can induce a magnetic field as described by one of Maxwell's fundamental equations: $\nabla \times B = \mu \left( J + \epsilon_0 \frac{\partial E}{\partial t} \right)$. The Ampere-Maxwell law states that a charged particle moving in a linear path will create a magnetic field perpendicular to the motion. That field can be amplified if, instead of linear, the charged particles move in a circular path where the magnetic fields that are perpendicular to the plane of motion add together. There are many ways to cause charged particles to move in a circular path, such as simply applying electrical current through curled wires in a solenoid. Another method is to utilize circularly polarized light to drive the circular motion of charged particles. A well-known example of this magneto-optical phenomenon is the inverse Faraday effect, which is when circularly polarized light excites the electrons in a material to move in a circular path, and thus generates a transient magnetic field when the driving light is present [1,2]. A more novel demonstration of this magneto-optical phenomenon is the phono-magnetic effect [3-7]. When circularly polarized light is directed into a material to drive ionic motion, the ions inside can gain angular momentum and rotate in a unidirectional path (so-called ion loops). In the most straightforward case, the ion loops are comprised of a pair of degenerate vibrational modes with orthogonal motion (E-symmetry modes in the material we study here). To achieve circular motion, the pair of degenerate modes must be excited with the correct relative phase delay. Note that at times these excited ion loops have been termed "chiral phonons", but simple 2D motion lacks the asymmetry as chiral excitations include a third dimension or propagation direction. Chiral phonons naturally break time-reversal symmetry (TRS), allowing materials to exhibit unique quantum phenomena, such as quantum Hall effect, superconductivity and magneto-electric effects as mentioned above [6,8-10].

This unique property of TRS breaking makes chiral phonons extremely useful in many novel applications, such as superconductors, spintronics, topological materials, and the focus of this paper – magnetism [11,12]. Achieving control of magnetic properties has great potential in data storage for electronic devices as a means to create ultrafast nonvolatile switches. Here we demonstrate the detection of an ultrafast Faraday rotation signal through the excitation of chiral phonon-polaritons in $LiNbO_3$, which is a demonstration of TRS breaking, and may additionally indicate the creation of a strong magnetic field that turns on faster than any switch currently used in electro-magnetic devices.

Phonon-magnetism via the excitation of ion loops has been experimentally observed in two recent publications. Luo, et. al. induced a Faraday rotation of ultrafast probe pulses in paramagnetic $CeF_3$ through phonon-spin coupling driven by circularly polarized THz light pulses [6]. Basini, et al. on the other hand, created a phonon-induced magnetooptical Kerr rotation of ultrafast probe pulses in diamagnetic $SrTiO_3$ (STO) [7]. STO has no innate magnetic ordering, meaning that they were able to switch on a magnetic-field driven Kerr signal in a non-magnetic material. Interestingly, Basini, et al. measured a Kerr rotation four orders of magnitude larger than had been theoretically calculated [3,5,13]. Whether or not the optical signals actually indicate such a strong magnetic field is under debate. Recently, Merlin theorized that in fact, the magnetic fields induced by both Luo, et. al and Basini, et al. were non-Maxwellian fields due to the breaking of TRS caused by the circularly polarized THz, but not true magnetic fields [6,7,14]. Merlin predicts that the magnitude difference between theoretical and experimental data in these experiments is due to the absence of the creation of a true magnetic field. In this work, we share some evidence that effects other than a magnetic field can result in an ultrafast Faraday signal; this may complicate optical detection of an induced magnetic response.

The magneto-optic Faraday effect is observed by the rotation of linearly polarized laser light in the presence of a magnetic field oriented in the light propagation direction. Here we experimentally observe the creation of a strong Faraday rotation resulting from phonon-magnetic excitation with circularly polarized THz pulses, similar to Refs. [6,7]. However, our sample and experimentation differ from these previous works in novel and important ways: firstly, the $LiNbO_3$ sample is a ferroelectric material where E-symmetry modes couple with light to form phonon-polaritons that propagate into the depth of the material [15], creating truly chiral ionic motion. Secondly, we can experimentally isolate Faraday signals that arise from linear polarization components of the chiral THz pump fields and remove these contributions from the measured response, in a way not possible in previous measurements.

Below we describe how we form chiral THz pulses to excite chiral phonon-polaritons in $LiNbO_3$, we show how to isolate the ultrafast Faraday rotation signal due to chiral excitation and separate it from signal arising from linearly polarized THz pulses, we model the signal contributions coming from both the inverse Faraday (electronic) effect and the phono-magnetic effect, and we show that this Faraday rotation is commensurate with the creation of a >10 T magnetic field in originally non-magnetic, ferroelectric $LiNbO_3$. However, we also shed light on the idea that the Faraday signal we observe does not solely arise from a magnetic response, leading to some uncertainty in the large magnetic field strength that we and others report.

**Experimental**

We utilized the two-dimensional (2D) THz experimental setup shown in Fig. 1 and described in Ref. [16] and Ref. [17]. The (1450 nm) and idler (1790 nm) outputs from our optical parametric amplifier (OPA) were directed into separate DAST THz generation crystals to generate broadband THz pulses with frequency content ranging from 0.5 to 5 THz (THz-1 and THz-2 respectively). THz-1 was vertically polarized and THz-2 was horizontally polarized, leading to the creation of perpendicularly polarized THz pulses to selectively excite the degenerate E modes. The two THz pump pulses were directed into a three-parabolic mirror scheme and focused onto the same area of the 500 $\mu$m thick, z-cut $LiNbO_3$ with a spot size of ~250 μm. By changing the relative timing between the THz pulses using the delay stage in the THz-2 beam path, we control the ellipticity of the combined THz light that irradiates the sample.

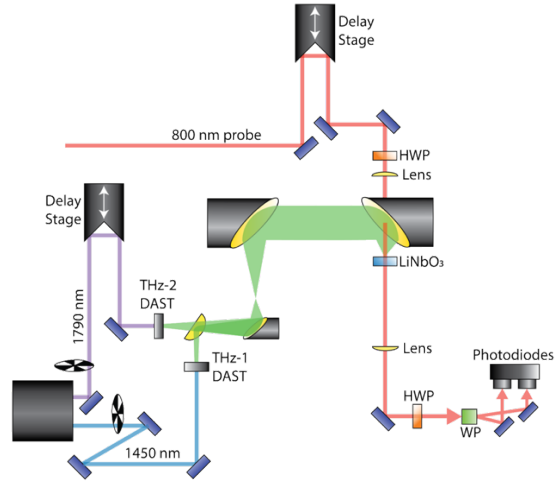

**Figure 1**. 2D THz experimental setup for chiral excitation. A pair of perpendicularly polarized THz pump beams (green) were created using ultrafast pulses of IR-light (purple and blue) and then focused onto the sample using a series of three parabolic mirrors. Delay between the THz pulses can be changed using the delay stage in the IR-pump path to control the ellipticity of the combined THz that excites the sample. The 800-nm probe beam was also focused onto the sample. Half-wave plates (HWP) in the probe path control the linear polarization of the Faraday probe and ensure the probe is polarized at 45 degrees at the Wollaston polarizer (WP) (in the absence of a sample response). The WP separates horizontal and vertical probe components to measure the polarization rotation with balanced photodiodes as a function of time with respect to the THz excitation.

We removed the signal contributions of linearly polarized THz pulses using a dual chopping scheme. In the THz-1 pump path, we placed a chopper spinning at 500 Hz, which blocks every other laser pulse at the native 1000 Hz repetition rate from the OPA. In the THz-2 path, we placed a chopper spinning at 250 Hz, which alternatively blocks and passes pairs of pulses at a time. This combination gave us access to the following pattern of THz pulses interacting with the 800 nm probe: both THz pulses, THz-1 only, THz-2 only, and no THz pulses. To isolate the nonlinear (chiral) signal, we subtract the THz-1 and THz-2 pulse responses from the response containing both THz pulses (see Eq. 1). We also subtracted the measured signal from the probe alone recorded when both THz pulses are blocked from each of the THz pulse-sequence responses (Eq. 1). As mentioned above, this nonlinear (chiral) signal isolation allows us to eliminate artifacts arising from excitation from the linearly polarized components of the chiral THz pump.

$$S_{nonlinear} = (S_{both} - S_{none}) - (S_{THz-1} - S_{none}) - (S_{THz-2} - S_{none}) \qquad \text{Eq. 1}$$

$S_{nonlinear}$, as detected in the polarization sensitive Faraday rotation scheme described in Fig. 1, measures the induced magnetic moment (oriented in the same or opposite direction to the probe propagation) as a function of time. Due to the Faraday effect, the polarization of a linearly polarized incident probe is rotated clockwise or counterclockwise around the direction of probe propagation when it passes through a medium influenced by a magnetic field. The magnitude of the polarization rotation depends linearly on the induced magnetic field ($M$), the material-dependent Verdet constant ($v$), and the thickness ($L$) of the medium, as shown in Eq. 2.

$$\Delta\theta = vML \qquad \text{Eq. 2}$$

We spatially overlapped a focused 800-nm probe beam with the THz on the LiNbO$_3$ sample (see Fig. 1). Using the first HWP we rotated the probe beam to a polarization of 45° before striking the sample. The second HWP ensured the probe beam was oriented such that an equal amount of vertically and horizontally polarized probe light struck the Wollaston prism (WP), and we measured a change in polarization of the linear probe beam through balanced detection with dual photodiodes. We varied the probe delay to record a series of scans over a probe range of ~ 4 ps. For each of several scans, the relative delay between THz pulses was varied by step sizes equivalent to a 6.67 fs delay on the THz-2 delay stage, which allowed us to measure the nonlinear signal at varying THz ellipticity angles.

With broadband circularly polarized THz pulses incident on a $z$-cut LiNbO$_3$, we directly excite all along the phonon-polariton dispersion curve of perpendicularly polarized E(TO$_1$) branches. In contrast to excitation in other materials of zone-center phonons with zero group velocity, excitation in LiNbO$_3$ results in chiral molecular vibrations (chiral ionic motions) propagating into the material with the main spectral amplitude ranging between 0 and 5 THz. Fig. 2 shows the lowest three E-phonon-polariton branches (blue lines) and our THz excitation spectrum in green.

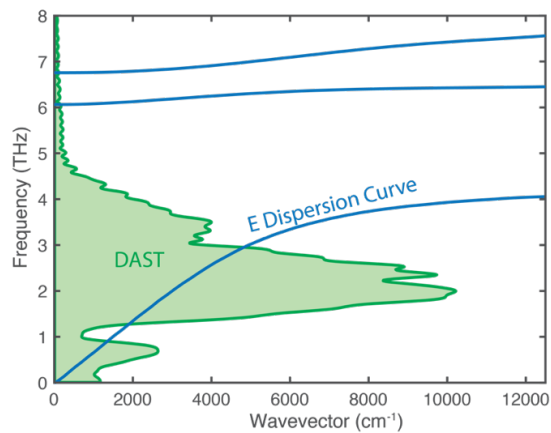

**Figure 2**. Green shading shows a representative spectrum from the DAST THz generation crystal used in these experiments, which THz can excite along the lowest of three phonon-polariton E mode branches shown by blue lines (dispersion curve extracted from Ref. [18]).

In addition to selective excitation of the $E(TO_1)$ branches to produce a chiral phonon-polariton that would induce a magnetic field, we also isolate the Faraday rotation from signal contributions due to the linear polarization components of the chiral THz pump field. Multiple processes exist that can rotate the polarization of an ultrafast probe as is measured in Kerr or Faraday rotation measurements. The previous experiments of Luo and Basini [6,7] do not isolate the Faraday signal from this linear pump-induced probe rotation. The differential chopping scheme we use allows us to isolate the Faraday signal from any probe rotations induced by linear components of the chiral THz excitation field.

**Results and Discussion**

To isolate the nonlinear (chiral) signal that is driven solely by the circular component of the excitation, we must remove the signal that arises from linear components of the THz pulses. As shown in Figure 3, we record signal with four pulse sequences that are separated with differential chopping. Fig. 3(a) shows the measured signal when only the vertically polarized THz-1 pulse excites the sample. We see clearly that linear (not circular) excitation results in the probe polarization being rotated – which does not indicate a TRS breaking magnetic-field driven Faraday signal. Presumably, these signals arise from the 800-nm probe interacting with the linearly polarized E phonon-polariton through a Raman process, which rotates the polarization of scattered probe light [15]. Fig. 3(b) shows a similar polarization rotation induced by the linearly polarized horizontal phonon-polariton motion driven by the THz-2 pump. Fig. 3(c) shows the sample response when the combined circularly polarized THz pulse excites the sample, showing that the signals from the linearly polarized V and H field components are present in the signal from the circularly polarized THz pulse. This combined signal in Fig. 3(c) shows artifacts from the linear components of the circular THz pump and is equivalent to the signals reported by Luo and Basini [6,7]. However, this measurement scheme can remove these artifact signals from the combined $LiNbO_3$ response by subtracting the linear pump signals shown in Fig. 3(a-b) from the combined signal in Fig.3(c) to isolate the Faraday rotation signal that arises purely due to the circularly polarized THz pump (Fig. 3(d)). The shape of the isolated nonlinear signal is dramatically different than the shape of the other three isolated pulses because we eliminate all the artifacts in the signal that arise from the Raman-scattering probe rotation caused by the oscillation of the phonon-polariton itself. The probe rotation isolated in the nonlinear signal Fig. 3(d) therefore purely arises from the chiral phonon-polariton excitation. However, the fact that a linearly polarized pump can induce a probe polarization rotation suggests that we must be careful in ascribing even our pure chiral pump signal to a magnetic Faraday response. In the following, we discuss the chiral signal as being a signature of an induced magnetic field, and below we revisit other possibilities.

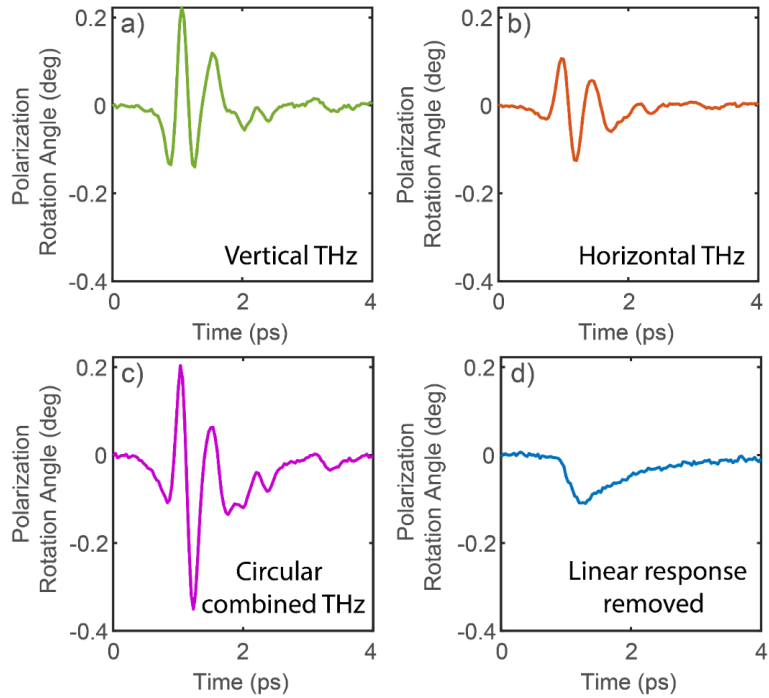

**Figure 3**. The pattern of four pulses isolated with the differential chopping scheme from measurements with two perpendicularly polarized THz pump pulses. a) and b) The signals from the vertical THz-1 pump and the horizontal THz-2 pump, respectively. c) The signal when both THz pulses are present to produce a chiral THz excitation pulse, showing clear contamination due to linear field response shown in (a) and (b). d) The isolated Faraday rotation signal that removes linear field components shown in (a) and (b).

In Fig. 4(a), we show the nonlinear (chiral-pump driven) Faraday rotation at the three most circular ellipticity angles of the THz pump: -118° (Fig. 4(c)), -36° (Fig. 4(d)), and 52° (Fig. 4(e)) (see Supplemental Materials for how the "average" ellipticity angle was determined).

Figure 4(a) shows the isolated Faraday rotation arises on an ultrafast, picosecond time scale. The Faraday rotation changes sign when switching from right-handed-circularly polarized (RHCP) THz to left-handed-circularly polarized (LHCP) input THz light, as expected for a signal arising from the inverse Faraday effect and the circular motion of ions. The change from right- to left-handed polarization reverses the order that the two phonon-polariton E modes are excited, changing the direction of ionic rotation, and reversing the direction of an induced magnetic field. In Fig. 4(b), we show how the magnitude of the Faraday signal changes with ellipticity angle, once again noting that the highest signals occur when the input THz pulse is more circular and essentially zero signal when the THz polarization and induced ionic motion is linear (0 degrees ellipticity angle). In Fig. 4(b), we plot the Faraday rotation signal amplitude against the ellipticity angles and compare to a model of the ellipticity angle signal dependence (discussed in more detail below).

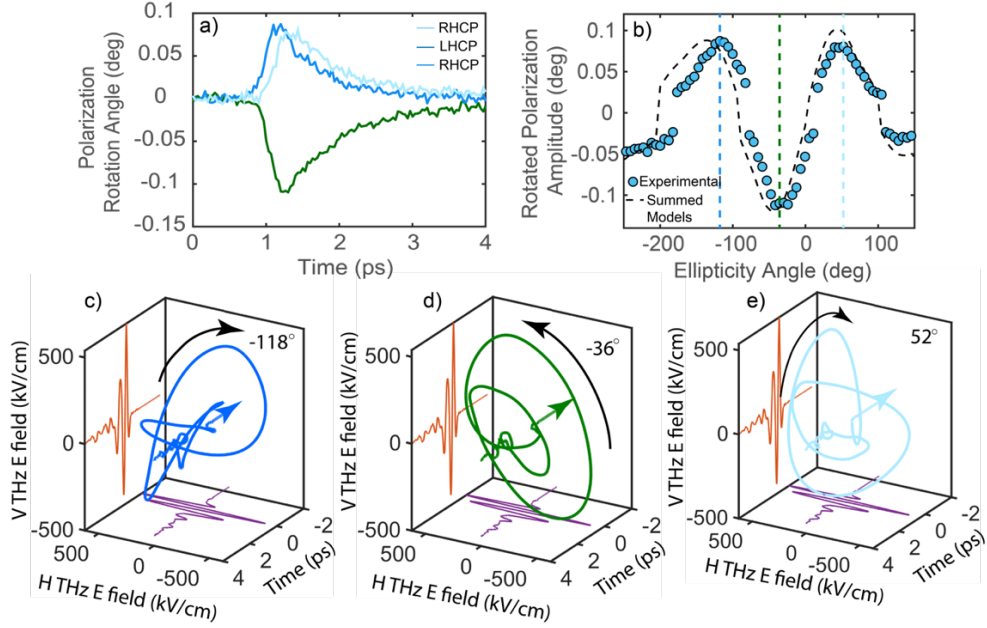

**Figure 4.** a) Isolated Faraday nonlinear signal arising from rotation of the probe polarization at the three most circular ellipticity angles. Dark blue is the response from left-handed circularly polarized (LHCP) input THz light and the lighter blues are the response from right-handed circularly polarized (RHCP) input THz light. b) The Faraday rotation signal magnitude as a function of ellipticity angle. The lines indicate the delays plotted in (a) The modeled signal is shown in black. (c)(d) and (e) show the combined motion of the THz pulses at each of the delays plotted in (a).

The isolated Faraday rotation signal follows the expected ellipticity angle dependence due to an induced magnetic field. However, in addition to the phono-magnetic effect and a magnetic field arising from the chiral phonon-polariton excitation, an induced magnetic field could also result from the inverse Faraday effect (IFE), where the circular motions of electrons in a material driven directly by the THz electric field create a magnetic field [19]. To determine whether a vibrational or electronic mechanism was creating the magnetic field that leads to the Faraday rotation in LiNbO$_3$, we mathematically modeled the response from both processes.

Based on Juraschek and co-workers' work [3], the magnetic moment of a charged particle is proportional to its angular momentum, and the angular momentum can be calculated from the motion of the charged particle as shown in Eq. 3. In Eq. 3, $M$ is the induced magnetic moment, $\gamma$ is a gyromagnetic ratio, $L$ is the angular momentum, while $Q_i$ and $\dot{Q}_i$ are the motion and velocity the charged species.

$$M = \gamma L = \gamma Q_i \times \dot{Q}_i \qquad \text{Eq. 3}$$

For the phononic contribution, $Q_i$ is the normal coordinate of individual phonon mode motion obtained previously at each point on the dispersion curve. We calculate a magnetic moment using Eq. 3 for each frequency of the phonon-dispersion curve and then sum up all the individual magnetic moments to obtain a total magnetic moment for the sample. Similarly, the electronic contribution is also in this form; however, now we need to consider the angular momentum of electrons instead, driven nonresonantly by the THz electric field $E_i$. Replacing $Q_i$ in Eq. 3 with the normal coordinate of the THz electric field enables us to model the IFE response ($M \propto E_i \times \dot{E}_i$) as well. To accurately account for the signal, we mathematically convoluted the signal with the probe pulse, which has a pulse duration of 100 fs.

We model $Q_i$ by computing the total motion of the phonon-polariton when excited with a broadband THz pulse by solving the differential equation of motion (Eq. 4) for each point on the dispersion curve (described in more detail in the Supplemental Material).

$$\ddot{Q}_i + 2\Gamma_i \dot{Q}_i + \omega_i^2 Q_i = Z_i^* E \qquad \text{Eq. 4}$$

In Eq. 4, $Q_i$ represents the vibrational coordinate of a particular part of the dispersion curve $i$, $\Gamma_i$ is the damping rate, $\omega_i$ is the angular frequency of the vibration, $Z_i^*$ is the mode effective charge, and $E$ is the THz driving field. We input the $\Gamma_i$ and $\omega_i$ values from Ref. [15] to calculate a separate $Q_i$ for each point on the dispersion curve. The $E$ used is an experimentally measured time trace of our THz pulse generated with DAST. By modeling in this matter, we generate an output of the vibrational mode motion ($Q_i$) at each frequency ($\omega_i$) that is excited by the THz bandwidth. We perform these calculations twice: once with the vertically polarized THz-1 field, and once with the variably delayed horizontally polarized THz-2 field to get the total motion.

In Fig. 5, we plot the magnetic field strength at the THz ellipticity angle (-34°) with the largest Faraday rotation. We observe a signal magnitude that would result from a peak magnetic field strength in excess of 10 Tesla, calculated using a Verdet constant of 0.2 °/T/cm from Ref. [20] (see Supplemental Materials for our own confirmation of the value). Figures 5(a) and 5(b) show that neither electronic or phononic modeled contributions respectively alone do not match the observed signal. Therefore, we linearly combine both contributions (black) as shown in Fig. 5(c). The linear combination, where only the relative magnitude of each is adjusted, agrees with the experimental data, suggesting that both phononic and electronic contributions lead to a magnetic field signal. To show this model represents well the signal from every THz ellipticity angle with the scaling factor from the fit at the -34° ellipticity angle, we combine the magnitude of the signal at each THz delay and sum the models together. Figure 4(b) shows this ellipticity-angle dependent model that agrees well with the data.

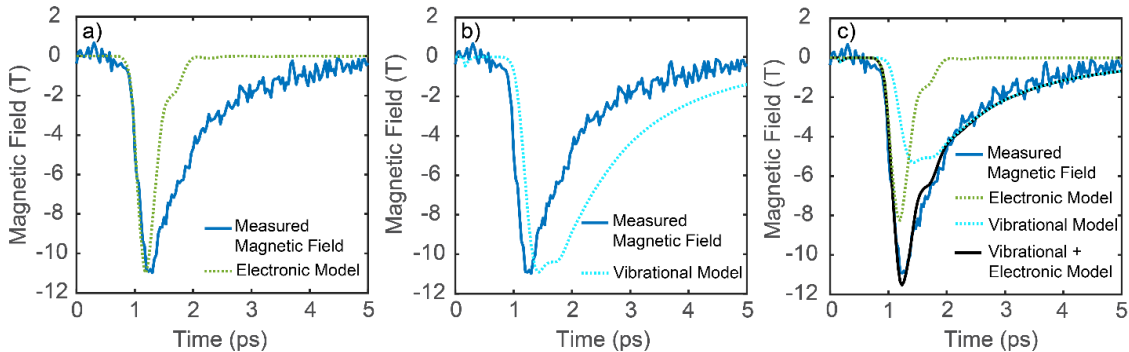

**Figure 5.** Experimental data of the induced magnetic field compared to the modeled magnetic field from a) the electronic (inverse Faraday effect) contribution, b) phononic contribution b) and the linear combination of both contributions (c). The y-axis magnitude is determined using the probe polarization rotation and the Verdet constant for $LiNbO_3$.

## Conclusion

In conclusion, we have excited ultrafast TRS-breaking chiral motion of phonon-polaritons. This leads to a magnetic field in the nonmagnetic material, LiNbO$_3$ that we attempt to quantify with Faraday rotation probe measurements where we can successfully eliminate other artifacts from the signal with differential chopping of the chiral THz pump. We show that the Faraday signal has a maximum value when the input combined THz pulse is circular, as predicted. The direction of the Faraday signal flips when the input THz pulse is changed from left- to right-circular polarization. We estimate a magnetic field strength of ~11 Tesla and determine through the fitting of mathematical models that the created magnetic field arises from both a vibrational and an electronic mechanism.

However, as discussed in the introduction, Merlin suggested the measured Faraday rotation in other experiments that had been ascribed to a phono-magnetism effect [6,7], may in fact be caused by the creation of non-Maxwellian fields. The fact that we see a Faraday rotation of our probe beam when excited with linearly polarized THz pulses suggests that a magnetic field isn't the only origin of the signal we measure. We propose that to differentiate the cause of the Faraday rotation, experiments must be performed with a variety of probes that will be sensitive to an induce magnetic field on ultrafast timescales. This may include x-ray probes of magnetic order, Zeeman splitting of appropriate states in adjacent materials that would be influenced by induced magnetic field lines, or potentially the modification of inherent internal magnetic order in some material with degenerate phonon modes like LiNbO$_3$ that would demonstrate the creation of a magnetic field to support or require a re-thinking of the optical probe measurements to date.

# Supplemental Material for: Faraday Rotation of Chiral Phonon-Polariton in LiNbO$_3$

## S1. Verdet Constant Measurements

We experimentally confirmed the Verdet constant from Ref. [20] through Faraday rotation measurements (see Fig. S1). In these measurements, we send the horizontally polarized probe beam through a Wollaston prism, splitting the beam into a large horizontal and a small vertical component which we direct into two different photodiodes. To create a static magnetic field, we place a pair of neodymium magnets (1.32 T) to the front and backside of a LiNbO$_3$ sample. The magnet has a hole in the center, which allows the probe beam to pass through the magnet unfettered and then transmits through the LiNbO$_3$. We record the magnitude of light hitting both photodiodes in five scenarios: 1) probe beam blocked, 2) probe beam unblocked with no LiNbO$_3$ or magnet, 3) probe beam unblocked with only the LiNbO$_3$ in the path, 4) probe beam unblocked with only the magnet in the path, and 5) probe beam unblocked with both the magnet and LiNbO$_3$ in the path. We subtract the reference with the light blocked from the other measurements. The magnet alone measurements assure us that the magnet itself is not blocking any probe amplitude because we get the same signal level with and without the magnet in, if there is no LiNbO$_3$. We then calculate the Faraday rotation angle based on the difference between the LiNbO$_3$ alone measurements and the LiNbO$_3$ + magnet measurements.

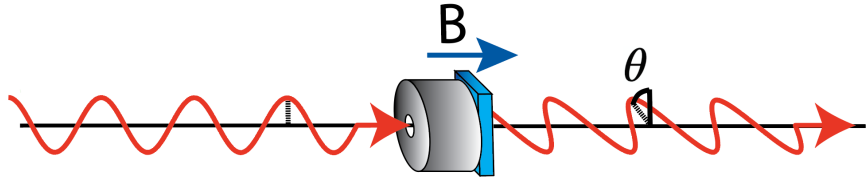

**Figure S1.** Faraday rotation measurements of the Verdet constant of LiNbO$_3$. The linearly polarized probe (red) rotates through the Faraday effect in the presence of a static magnetic field (caused by the gray magnet), which is directed through a Wollaston prism and into two photodiodes.

To calculate the Verdet constant ($v$) we use the typical Faraday rotation equation shown below (Eq. S1). We have rearranged the equation to solve for $v$, where L is the thickness of the LiNbO$_3$ (500 µm), B is the strength of the magnetic field applied to the crystal (0.537 T, calculated based on our experimental setup), and $\Delta\theta$ is the experimental change in polarization between the LiNbO$_3$ alone and the LiNbO$_3$ + magnet measurement. We only take into account the change in the small amount of vertically polarized light as we are more sensitive to a change in a small number than a small change in a large number.

$$v = \Delta\theta BL \qquad \text{Eq. S1}$$

We measured a Verdet constant comparable to the Verdet constant from Ref. [20].

## S2. Modeling the Phonon-Polariton Motion

We modeled the total motion of the phonon-polariton when excited with a broadband THz pulse. We began by manually extracting the phonon-polariton dispersion curve and damping rates from

Fig. 5 in Ref. [15]. For the dispersion curve, we performed a fit on the extracted values to create an equation for the curve (Eq. S2) (plotted in Fig. S2). Using Eq. S2, we calculate frequencies for 500 evenly spaced points on the dispersion curve between 1000 cm$^{-1}$ and 20000 cm$^{-1}$. The odd shape of the damping rate limited our ability to accurately fit it with an equation. Instead, we splined the extracted values from Ref. [15] and in this case created 500 data points evenly spaced between 0 cm$^{-1}$ and 20000 cm$^{-1}$ plotted in Fig. S2.

$$\omega_i = 4\pi(-3.747 \times 10^{-17}x^4 + 2.937 \times 10^{-12}x^3 - 8.142 \times 10^{-8}x^2 + 9.723 \times 10^{-4}x - 0.1832) \quad \text{Eq. S2}$$

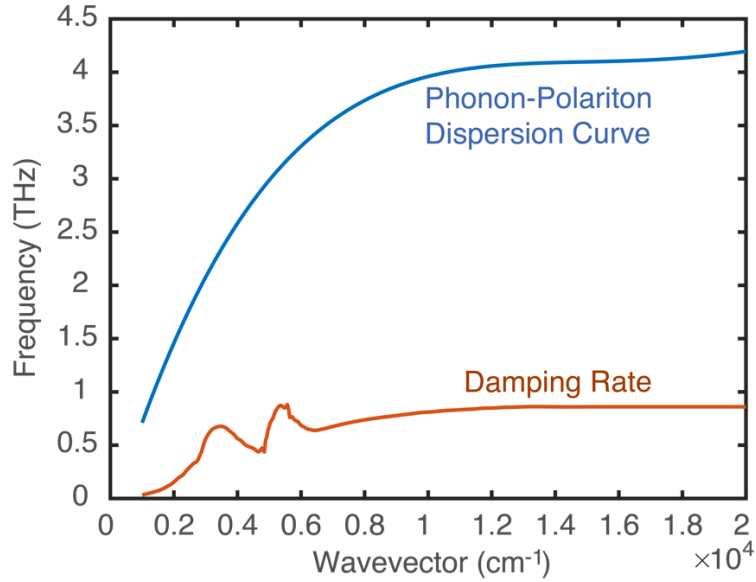

**Figure S2**. The E mode phonon-dispersion curve (blue) and damping rate (red) for LiNbO$_3$ extracted from Ref. [15].

We then solve the differential equation of motion (Eq. S3) for each point on the dispersion curve.

$$\ddot{Q}_i + 2\Gamma_i \dot{Q}_i + \omega_i^2 Q_i = Z_i^* E \quad \text{Eq. S3}$$

In Eq. S3, $Q_i$ represents the vibrational coordinate of a particular part of the dispersion curve $i$, $\Gamma_i$ is the damping rate, $\omega_i$ is the angular frequency, $Z_i^*$ is the mode effective charge, and $E$ is the electric field. We input the $\Gamma_i$ and $\omega_i$ values from Fig. S2, and we calculate a separate $Q_i(t)$ for each of 500 discrete points along the dispersion curve. The THz electric field $E$ used is an experimentally measured time trace of our THz pulse generated with DAST. $Z_i^*$ was an arbitrary constant value kept the same for each point on the dispersion curve. This outputs a model of the vibrational mode motion at each frequency in arbitrary units. Fig. S3(a) and S3(b) show the vibrational motion and its Fourier transform at 4 different points along the dispersion curve near 1, 2, 3, and 4 THz. The damping rate is lower at low frequencies so the oscillations at the lower frequencies last longer in time, while the high frequency oscillations damp out quickly. The amplitude of the oscillations relates to the THz amplitude being input at the specific frequencies of the oscillations. We model the vibrational coordinate for both perpendicular branches of the phonon-dispersion curve through the same process, but change the experimental THz pulse that is read into the differential equation from the vertical THz-1 trace to the horizontal THz-2 trace.

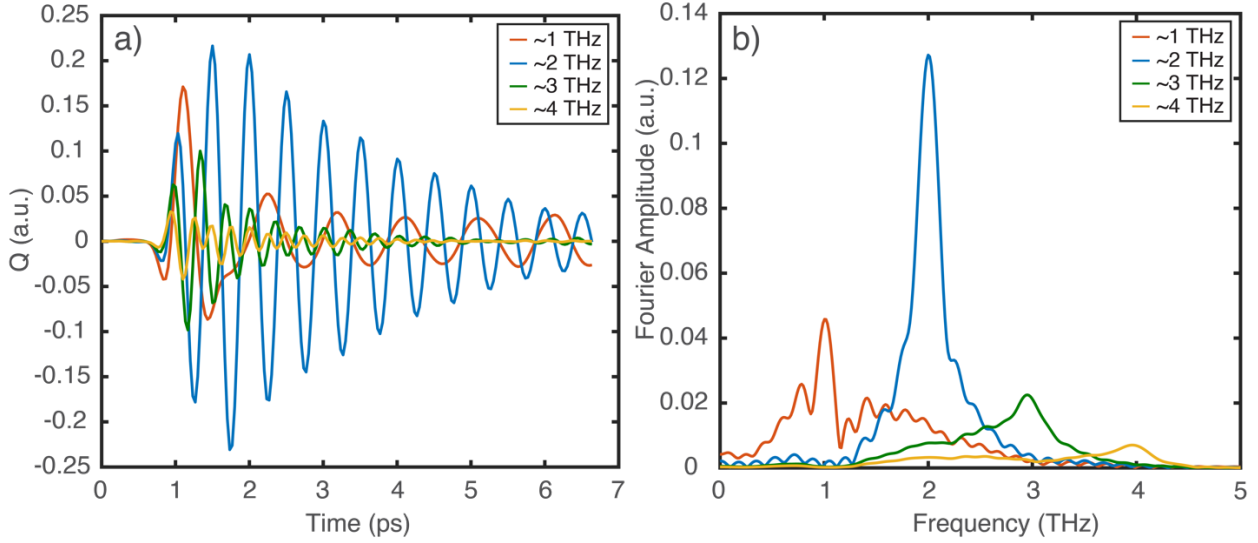

**Figure S3**. A) The modeled vibrational coordinate of the E phonon-polariton at four frequencies on the dispersion curve and b) the Fourier transform of those vibrational coordinates.

The calculated $Q_i$ values along the dispersion curve are then used to calculate the magnetic moment that arise from individual vibration as shown in Eq. 3. To simplify the process, we treat the gyromagnetic ratio as unity and look at the time-resolved shape of the response. The strength of the overall magnetic field is equal to the sum of $M_i$ at all frequencies together for the entire phonon dispersion curve. We perform this calculation at every delay between the THz pulses. Fig. S4(a) shows the individual magnetic field models calculated at 4 frequencies along the phonon dispersion curve. Fig. S5(b) shows the totals sum of those 4 modeled magnetic moments. These are the modeled magnetic moments only at the delay between THz pulses where the largest Faraday signal was isolated (at an ellipticity angle of -34°).

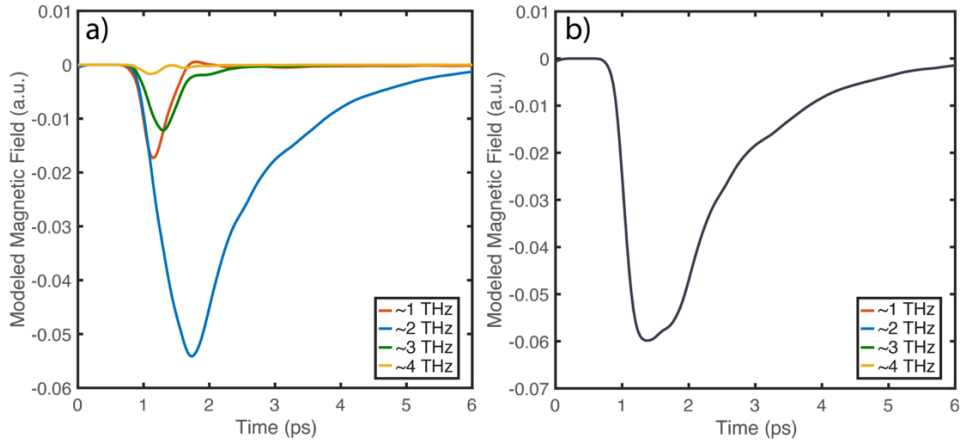

**Figure S4**. (a) Modeled magnetic field from vibrational motion of the E(TO$_1$) phonon-polariton at 4 frequencies. (b) Total modeled magnetic field from summing the magnetic field models for 4 frequencies.

We modeled the magnetic field induced by electronic contribution with a very similar approach – instead of treating $Q_i$ as vibrational coordinate, it is treated as the THz-electric-field-driven electron oscillation.

## S3. Calculating the Ellipticity Angle

We calculate the ellipticity angle of the combination of the horizontal and vertical THz pulse at each delay between pulses. Determining true ellipticity angles for a few-cycle pulse is not exact (see Ref. [21]), so we calculate average ellipticity angle directly from the delay between the THz pulses. We chose to use DAST THz generation crystals for these experiments because the two-cycle THz pulses simplify determination of an ellipticity angle, compared to a more efficient THz generator like PNPA [22], which generates approximately single-cycle THz pulses.

Before every set of sample measurements, we use electro-optic sampling to measure the vertical and horizontal THz pulses at a single delay. In post, we shifted the initial horizontal pulse experimental measurement to the correct delay for each stage positions in the LiNbO$_3$ measurements based on the difference in stage position of this initial measurement and the stage positions of the subsequent measurements on the LiNbO$_3$ sample. From the subsequent shifted horizontal pulse scans and the initial vertical pulse scan, we calculate the ellipticity angle using three different methods, all based on Eq. S5.

$$\frac{\Delta x}{2\lambda} = \text{ellipticity angle} \quad \quad \text{Eq. S5}$$

Where $\lambda$ is the wavelength of the horizontal THz pulse and $\Delta x$ is the averaged distance between peaks in the horizontal and vertical THz pulses. We use three different methods to calculate $\lambda$, shown in Fig. S5. For Method 1 (Fig. S5(a)), we assume $\lambda$ is the distance between the first two maxima. For Method 2 (Fig. S5(b)), we calculate as twice the distance between the first large maxima and the largest minima. For Method 3 (Fig. S5(c)), $\lambda$ is determined as twice the distance between the largest minima and the second maxima.

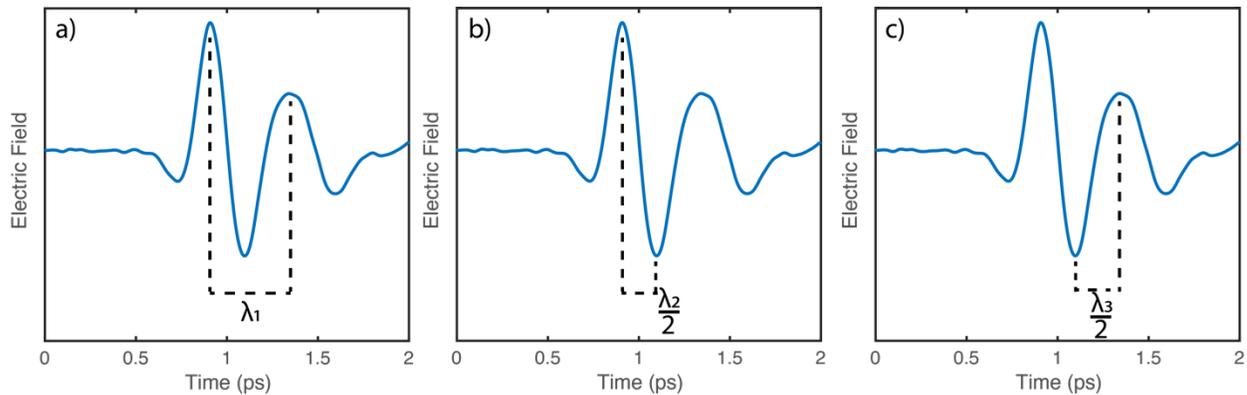

**Figure S5. Calculating the wavelength of the THz electric field strength in three ways.** a) Method 1: The distance between the first two minima. b) Method 2: Twice the distance between the first maxima and the first minima. c) Method 3: Twice the distance between the first minima and the second maxima.

We then calculate $\Delta x$ at each stage position based on the delay between the horizontal and vertical THz pulses. We calculate a separate $\Delta x$ for each major peak in the THz traces (shown in Fig. S6) and then average the three calculated $\Delta x$s to get a single value for $\Delta x$ to use in Eq. S5.

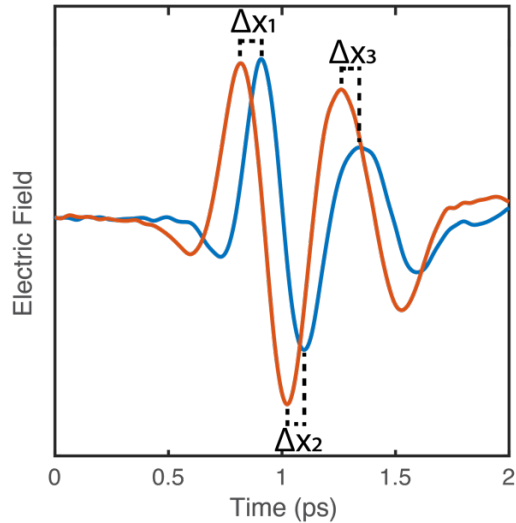

**Figure S6**. Illustration of how the individual $\Delta x$s are calculated.

For the delay shown in Fig. S6, which is also the delay in the main paper referred to as LHCP, these three methods were used and calculated ellipticity angles of -36°, -33°, and -31°. The ellipticity angle used in Fig. 4 were calculated with Method 2.

We note that the most circular polarization of the combined THz pulses should occur at 45° and -45°, so our calculation of the ellipticity angle is imperfect. This can likely be attributed to the fact that the two input THz pulses are generated using different DAST THz generation crystals, and therefore the shapes of the waveforms are not identical, affecting the ellipticity angle calculation. It may also be that the sample position in the focus of the EO crystal when measuring the THz pulses differs slightly from the LiNbO$_3$ position in the focus, slightly changing the delay between pulses and leading to a small offset between experiment and theory in Figure 4(b).